\documentclass{sig-alternate}

\usepackage[english]{babel}
\usepackage{xspace}
\usepackage{algorithm}
\usepackage[noend]{algorithmic}

\usepackage{graphicx}

\newcommand{\calO} {\mathcal O}
\newcommand{\cE} {\mathcal E}
\newcommand{\cG} {\mathcal G}
\newcommand{\cV} {\mathcal V}
\newcommand{\frakg} {\mathfrak g}

\newcommand{\Name}[1]{{\emph{#1}}\xspace}

\newcommand{\menor}{\leqslant} 
\newcommand{\vmenos}{\vspace{-0.3cm}}

\newcommand{\Figure}[5]{
\begin{figure}[#2]
\centering
\includegraphics[width=#3]{#1}
\vspace{-0.5cm}
\caption{#5}
\label{fig:#4}
\end{figure}
}

\newtheorem{lemma}{Lemma}
\newtheorem{proposition}{Proposition}

\newdef{problem}{Problem}

\begin{document}
\conferenceinfo{AISec'11,} {October 21, 2011, Chicago, Illinois, USA.}
\CopyrightYear{2011}
\crdata{978-1-4503-1003-1/11/10}
\clubpenalty=10000
\widowpenalty = 10000

\title{An Algorithm to Find Optimal Attack Paths in Nondeterministic Scenarios}

\numberofauthors{3}

\author{
\alignauthor
Carlos Sarraute\\
       \affaddr{Core Security Technologies and ITBA}\\
       \affaddr{Buenos Aires, Argentina}\\
       \email{carlos@corest.com}
\alignauthor
Gerardo Richarte\\
       \affaddr{Core Security Technologies}\\
       \affaddr{Buenos Aires, Argentina}\\
       \email{gera@corest.com}
\alignauthor
Jorge Luc\'angeli Obes\\
       \affaddr{Universidad de Buenos Aires}\\
       \affaddr{Argentina}\\
       \email{jlucangeli@dc.uba.ar}
}

\date{30 June 2011}

\maketitle

\begin{abstract}

As penetration testing frameworks have evolved and have become more complex, 
the problem of controlling automatically the pentesting tool
has become an important question.
This can be naturally addressed as an \emph{attack planning} problem.
Previous approaches to this problem were based
on modeling the actions and assets in the PDDL language, and using off-the-shelf AI tools to generate attack plans. These approaches however are limited. In particular, the planning is classical (the actions are deterministic) and thus not able to handle the uncertainty involved in this form of attack planning. 
We herein contribute a planning model that does capture the uncertainty about the results of the actions, which is modeled as a probability of success of each action. We present efficient planning algorithms, specifically designed for this problem, that achieve industrial-scale runtime performance
(able to solve scenarios with several hundred hosts and exploits). 
These algorithms take into account the probability of success of the actions and their expected cost (for example in terms of execution time, or network traffic generated).
We thus show that probabilistic attack planning can be solved efficiently
for the scenarios that arise when assessing the security of large networks.
Two ``primitives'' are presented, which are used as building blocks
in a framework separating the overall problem into two levels of abstraction.
We also present the experimental results obtained with our implementation,
and conclude with some ideas for further work.
\end{abstract}

\category{C.2.0}{Computer-Communication Networks}{General}[Security and protection]
\category{I.2.8}{Artificial Intelligence}{Problem Solving, Control Methods, and Search}[Plan execution, formation, and generation]
\category{K.6.5}{Management of Computing and Information Systems}{Security and Protection}[Unauthorized access]
\category{K.6.m}{Management of Computing and Information Systems}{Miscellaneous}[Security]

\terms{Security}

\keywords{Network security, exploit, automated pentesting, attack planning}

\section{Introduction}

Penetration testing has become one of the most trusted ways of assessing the
security of networks large and small. The result of a penetration test is a
repeatable set of steps that result in the compromise of particular assets in the network.
Penetration testing frameworks have been developed to
facilitate the work of penetration testers 
and make the assessment of network security more accessible to non-expert users \cite{BurKilBea07}.
The main tools available are the commercial products
Core Impact (since 2001), Immunity Canvas (since 2002),
and the open source project Metasploit
(launched in 2003, owned by Rapid7 since 2009).
These tools have the ability
to launch actual exploits for vulnerabilities, contributing to expose risk
by conducting an attack in the same way an external attacker would \cite{ArcGra04}.


As pentesting tools have evolved and have become more complex -- covering new
attack vectors, and shipping increasing numbers of exploits and information gathering techniques --
the problem of controlling the pentest framework successfully has become an important question.
A computer-generated plan for an attack would isolate the user from the
complexity of selecting suitable exploits for the hosts in the target network.
In addition, a suitable model to represent these attacks would help to
systematize the knowledge gained during manual penetration tests performed
by expert users, making pentesting frameworks more accessible to non-experts.



A natural way to address this issue is as an {\em attack planning} problem.
This problem was introduced to the AI planning community by Boddy \emph{et al.} 
as the ``Cyber Security'' domain \cite{BodGohHaiHar05}.
In the pentesting industry, Lucangeli \emph{et al.} proposed a solution based 
on modeling the actions and assets in the PDDL language,\footnote{PDDL stands for 
Planning Domain Definition Language. 
Refer to \cite{FoxLon03} for a specification of PDDL 2.1.}
and using off-the-shelf 
planners to generate attack plans \cite{LucSarRic10}. 
Herein we are concerned with the specific context of regular automated pentesting,
as in ``Core Insight Enterprise'' tool.
We will use the term ``attack planning'' in that sense.

Recently, a model based on partially observable Markov decision processes (POMDP)
was proposed, in part by one of the authors \cite{SarBufHof11}.
This grounded the attack planning problem in a well-researched formalism,
and provided a precise representation of the attacker's uncertainty
with respect to the target network. In particular, the information gathering phase
was modeled as an integral part of the planning problem.
However, as the authors show, this solution does not scale to medium or large real-life networks.

In this paper, we take a different direction: 
the uncertainty about the results of the actions is modeled 
as a {\em probability of success} of each action, 
whereas in \cite{SarBufHof11} the uncertainty is modeled
as a distribution of probabilities over the states.
This allows us to produce an efficient planning algorithm,
specifically designed for this problem,
that achieves industrial-scale runtime performance.

Of course planning in the probabilistic setting is far more difficult 
than in the deterministic one.
We do not propose a general algorithm, but a solution suited
for the scenarios that need to be solved in a real world penetration test.
The computational complexity of our planning solution is $\calO ( n \log n) $,
where $n$ is the total number of actions in the case of an attack tree
(with fixed source and target hosts),
and $\calO ( M^{2} \cdot n \log n) $ where $M$ is the number of machines
in the case of a network scenario.
With our implementation, we were able to solve planning
in scenarios with up to 1000 hosts distributed in different networks.

We start with a brief review of the attack model in Section \ref{sec:model},
then continue with a presentation of two ``primitives'' in 
Sections \ref{sec:choose} and \ref{sec:combine}.
These primitives are applied in more general settings in 
Sections \ref{sec:dynamic} and \ref{sec:distinguished-assets}.
Section \ref{sec:experiments} shows experimental results from the implementation
of these algorithms.
We conclude with some ideas for future work.

%
%
%

\section{The Attack Model}\label{sec:model}
We provide below some background on the 
conceptual model of computer attacks that we use,
for more details refer to \cite{ArcRic03,FutNotRic03,Richarte03,RusTis07}. 
This model is based on the concepts of assets, goals, agents and actions.
In this description, an attack involves a set of agents, 
executing sequences of actions, 
obtaining assets (which can be information
or actual modifications of the real network and systems) in order to reach a set of goals.

An \textit{asset} can represent anything that an attacker may need to
obtain during the course of an attack, including the actual goal. 
Examples of assets:
information about the Operating System (OS) of a host $H$;
TCP connectivity with host $H$ on port $P$;
an Agent installed on a given host $H$.
To \emph{install an agent} means to break into a host,
take control of its resources, 
and eventually use it as pivoting stone to continue the attack
by launching new actions based from that host.

The \textit{actions} are the basic steps which form an attack.
Actions have requirements (also called preconditions)
and a result: the asset that will be obtained if the action
is successful.
For example, consider the exploit 
\Name{IBM Tivoli Storage Manager Client Remote Buffer 
Overflow}\footnote{The particular implementations that we have studied
are the exploit modules for Core Impact and Core Insight Enterprise, although the same model
can be applied to other implementations, such as Metasploit.}
for the vulnerabilities in \emph{dsmagent} described by CVE-2008-4828.
The result of this action is to install an agent,
and it requires that the OS of the target host is 
Windows 2000, Windows XP, Solaris 10, Windows 2003, or AIX 5.3.
In this model, all the exploits (local, remote, client-side, webapps) are represented
as actions. Other examples of actions are:
TCP Network Discovery, UDP Port Scan, DCERPC OS Detection,
TCP Connectivity Probe.

The major differences between the attack model used in this work
and the {\em attack graphs} used in 
\cite{AmmWijKau02,JajNoeBer05,JhaSheWin02,PhiSwi98,RitAmm00,SheHaiJha02}
are twofold: 
to improve the realism of the model, we consider that
the actions can produce numerical effects (for example, the expected running time of each action);
and that the actions have a probability of success (which models the uncertainty about the
results of the action).

\subsection{Deterministic Actions with Numerical Effects}

In the deterministic case, the actions and assets that compose a specific planning
problem can be successfully represented in the PDDL language.
This idea was proposed in \cite{SarWei08} and further analyzed in \cite{LucSarRic10}.
The assets are represented as PDDL predicates, and the actions
are translated as PDDL operators.
The authors show how this PDDL representation allowed
them to integrate a penetration testing tool with an external planner,
and to generate attack plans in realistic scenarios.
The planners used -- Metric-FF \cite{Hoffmann02} and SGPlan \cite{CheWahHsu06} --
are state-of-the-art planners able to handle numerical effects.

Fig.~\ref{fig:pddl-action} shows an example of a PDDL action: an exploit for the 
IBM Tivoli vulnerability, that will attempt
to install an agent on target host $t$ from an agent previously installed
on the source host $s$. To be successful, this exploit requires that
the target runs a specific OS, has the service {\ttfamily mil-2045-47001} running and
listening on port 1581.


\begin{figure}[ht]
{
\footnotesize

\begin{verbatim} 
(:action IBM_Tivoli_Storage_Manager_Client_Exploit
:parameters (?s - host ?t - host)
:precondition (and
  (compromised ?s)
  (and (has_OS ?t Windows)
    (has_OS_edition ?t Professional)
    (has_OS_servicepack ?t Sp2)
    (has_OS_version ?t WinXp)
    (has_architecture ?t I386))
  (has_service ?t mil-2045-47001)
  (TCP_connectivity ?s ?t port1581)
)
:effect(and 
  (installed_agent ?t high_privileges)
  (increase (time) 4)
))
\end{verbatim} 
}
\vmenos
\caption{Exploit represented as PDDL action.}
\label{fig:pddl-action}
\end{figure}


The average running times of the exploits are measured by 
executing all the exploits of the penetration testing tool in a testing lab.
More specifically, in Core's testing lab there are more than 
748 virtual machines with different OS and installed applications,
where all the exploits of Core Impact are executed every night \cite{Pico06}.

\subsection{Actions' Costs}

The execution of an action has a multi-dimensional cost. We detail below
some values that can be measured (and optimized in an attack):

\begin{description}
	\item[Execution time:] Average running time of the action.

	\item[Network traffic:] The amount of traffic sent over the network
	increases the level of noise produced.

	\item[IDS detection:] Logs lines generated and alerts triggered by the execution of the 
	action increase the noise produced.

	\item[Host resources:] The execution of actions will consume resources of 
	both the local and remote host, in terms of CPU, RAM, hard disc usage, etc.

  \item[Traceability of the attack:] Depends on the number of intermediate hops and topological factors.

	\item[Zero-day exploits:] Exploits for vulnerabilities that are not
	publicly known are a valuable resource, that should be used only when 
	other exploits have failed (the attacker usually wants to minimize the use 
	of ``0-days'').
\end{description}

In our experiments, we have chosen to optimize the expected execution time.
In the context of regular penetration tests, 
minimizing the expectation of total execution time is a way of
maximizing the amount of exploits successfully launched in a fixed time frame
(pentests are normally executed in a bounded time period).

However, the same techniques can be applied to any other scalar cost,
for example to minimize the noise produced by the actions
(and the probability of being detected).

\subsection{Probabilistic Actions}

Another way to add realism to the attack model is to consider that the actions
are nondeterministic. This can be modeled by associating probabilities 
to the outcomes of the actions. In the case of an exploit,
the execution of the exploit can be successful (in that case the attacker
takes control of the target machine) or a failure. This is represented
by associating a \emph{probability of success} to each exploit.

The probability of success is conditional: it depends on the environment conditions.
For example, the IBM Tivoli exploit for CVE-2008-4828 is
more reliable (has a higher probability of success) 
if the OS is Solaris since it has no heap protection,
the stack is not randomized and is executable by default.
Alternatively, the exploit is less reliable (has a lower probability of success)
if the OS is Windows XP SP2 or Windows 2003 SP1, 
with Data Execution Prevention (DEP) enabled.
On Windows Vista, the addition of 
Address Space Layout Randomization (ASLR)
makes the development of an exploit even more difficult, and diminishes its 
probability of success.
In practice, the probability of success of each exploit is measured
by exhaustively executing the exploit against a series of targets,
covering a wide range of OS and application versions.

Although it improves the realism of the model, considering probabilistic
actions also makes the planning problem more difficult.
Using general purpose probabilistic planners did not work
as in the deterministic case;
for instance, we experimented with Probabilistic-FF \cite{DomHof07}
with poor results, since it was able to find plans in only very small cases.

In the rest of this paper, we will study algorithms
to find optimal attack paths in scenarios of increasing difficulty.
We first describe two primitives, and then apply them 
in the context of regular automated pentesting.
In these scenarios we make an additional hypothesis:
the independence of the actions. 
Relaxing this hypothesis is a subject for future work.

\section{The {\secit Choose} Primitive}\label{sec:choose}

We begin with the following basic problem.
Suppose that the attacker (i.e. pentester) wants to gain access to the credit cards 
stored in a database server $H$ by installing a system agent.
The attacker has a set of $n$ remote exploits that he can launch against that server.
These exploits result in the installation of a system agent when successful 
(see Fig.~\ref{fig:grafo1}).

\Figure{grafo1.pdf}{ht}{\linewidth}{grafo1}
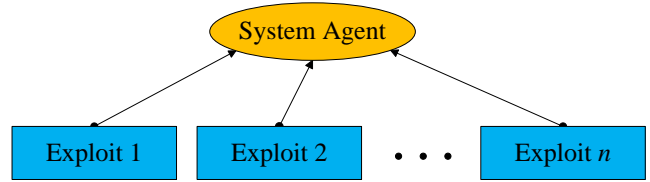
{Multiple exploits may install a System agent (on the target host).}

In this scenario, the attacker has already performed information gathering
about the server $H$, collecting a list of open/closed ports,
and running an OS detection action such as Nmap.
The pentesting tool used provides statistics on
the probability of success and expected running time for each exploit 
in the given conditions.\footnote{In our experiments
we used the database of tests of
Core Impact and Core Insight Enterprise.}
The attacker wants to minimize the expected execution time of the whole attack.
A more general formulation follows:

\begin{problem}\label{action_order}
Let $\frakg$ be a fixed goal, and let $\{ A_1, \ldots, A_n  \}$ be a set 
of $n$ independent actions whose result is $\frakg$.
Each action $A_k$ has a probability of success $p_k$ and expected
cost $t_k$. 
Actions are executed until an action is successful and provides the goal $\frakg$
(or all the actions fail).

\noindent
Task: Find the order in which the actions must be executed
in order to minimize the expected total cost.

\end{problem}

We make the simplifying assumption that the probability of success of each action 
is independent from the others.
If the actions are executed in the order $A_1, \ldots, A_n$,
using the notation $ \overline{p_i} = 1 - p_i $, the expected cost
can be written as
\begin{align}\label{exp_time_or}
T_{\{ 1 \ldots n \} } = t_1 + \overline{p_1} \, t_2 + \ldots 
+ \overline{p_1} \, \overline{p_2} \ldots \overline{p_{n-1}} \, t_n .
\end{align}
The probability of success is given by
$$
P_{ \{ 1 \ldots n \} } = p_1 + \overline{p_1} \, p_2 + 
\overline{p_1} \, \overline{p_2} \, p_3 + \ldots +
\overline{p_1} \ldots \overline{p_{n-1}} \, p_n ,
$$
and the complement 
$\overline{ P_{ \{ 1 \ldots n \} }}  
= \overline{p_1} \, \overline{p_2} \ldots \overline{p_{n}} $.
In particular this shows that the total probability of success does not depend on the order of
execution.

Even though this problem is very basic, we didn't find references to its solution. 
This is why we give below some details on the solution that we found.

\begin{lemma} \label{lemma-ordered-actions}
Let $A_1, \ldots, A_n$ be actions such that
$t_1 / p_1 \menor t_2 / p_2 \menor \ldots \menor t_n / p_n $.
Then
$$
\frac{ T_{ \{1 \ldots n-1 \} } } { P_{ \{1 \ldots n-1 \} } } \menor \frac{t_n}{p_n} .
$$
\end{lemma}

\begin{proof}
We prove it by induction. The case with two actions is trivial,
since we know by hypothesis that $t_1 / p_1 \menor t_2 / p_2 $.
For the inductive step, suppose that the proposition holds for $n-1$ actions.
Consider the first three actions $A_1, A_2, A_3$.
The inequality 
\begin{align*}
\frac{T_{\{12\}}}{P_{\{12\}}} \menor \frac{t_3}{p_3}
\end{align*}
holds if and only if
$ t_2 / p_2  \menor t_3 / p_3 $.
So the first two actions can be considered as a single action $A_{12}$ with
expected cost (e.g. running time) $T_{\{12\}}$ and probability of success $P_{\{12\}}$.
We have reduced to the case of $n-1$ actions, and we can use the induction hypothesis
to conclude the proof. 
\end{proof}

\begin{proposition}\label{prop_action_order}
A solution to Problem \ref{action_order} is to sort the actions according to the 
coefficient $t_k / p_k$ (in increasing order), and to execute them in that order.
The complexity of finding an optimal plan is thus $\calO ( n \log n ) $.
\end{proposition}

\begin{proof}
We prove it by induction. We begin with the case of
two actions $A_i$ and $A_j$ such that $t_i / p_i \menor t_j / p_j$.
It follows easily that $- p_i t_j \menor - p_j t_i $ and that
$$
t_i + (1 - p_i) \, t_j \menor t_j + (1 - p_j) \, t_i .
$$
For the inductive step, suppose for the moment
that the actions are numbered so that
$t_1 / p_1 \menor \ldots \menor t_n / p_n $,
and that the proposition holds for all sets of $n-1$ actions.
We have to prove that executing $A_1$ first is better
that executing any other action $A_k$ for all $k \neq 1$.
We want to show that
$$
\begin{array}{c}
t_1 + \displaystyle\sum_{2 \leqslant i \leqslant n} t_i \cdot 
\prod_{1 \leqslant j \leqslant i-1} \overline{p_j}  \\
 \menor \: \:
t_k + \displaystyle\sum_{1 \leqslant i \leqslant n, \, i \neq k} t_i \cdot 
\overline{p_k} \cdot 
\prod_{1 \leqslant j \leqslant i-1, \, j \neq k} \overline{p_j} .
\end{array}
$$
Notice that in the two previous sums, the coefficients of $t_{k+1}, \ldots, t_n$
are equal in both expressions.
They can be simplified,
and using notations previously introduced, the inequality can be rewritten
$$
T_{ \{1 \ldots k-1 \} } + \overline{ P_{ \{1 \ldots k-1 \} } } \; t_k 
 \menor t_k + \overline{p_k} \; T_{ \{1 \ldots k-1 \} }
$$
which holds if and only if
$$
\frac{ T_{ \{1 \ldots k-1 \} } } { P_{ \{1 \ldots k-1 \} } }  \menor  \frac{t_k}{p_k}
$$
which is true by Lemma \ref{lemma-ordered-actions}.
We have reduced the problem to sorting the coefficients $t_k / p_k$.
The complexity is that of making the $n$ divisions $t_k / p_k$
and sorting the coefficients.
Thus it is $\calO ( n + n \log n ) = \calO ( n \log n )$.
\end{proof}

We call this the {\em choose} primitive because it tells you,
given a set of actions, which action to choose first: 
the one that has the smallest $t/p$ value.
In particular, it says that you should execute first the actions
with smaller cost (e.g. runtime) or higher probability of success,
and precisely which is the trade-off between these two dimensions.

The problem of choosing the order of execution within a set of exploits
is very common in practice. 
In spite of that, 
the automation methods currently implemented in penetration testing frameworks
offer an incomplete solution,\footnote{As of July 2011, Immunity Canvas \cite{Aitel04}
doesn't provide automated execution of exploits; Metasploit \cite{Moore10} has
a feature called {\em ``autopwn''} that launches all the exploits available 
for the target ports in arbitrary order; Core Impact Pro launches
first a set of {\em ``fast''} exploits and then {\em ``brute-force''} exploits \cite{SarWei08}, but
arbitrary order is used within each set; Core Insight Enterprise
uses planning techniques based on a PDDL description \cite{LucSarRic10} that takes into account
the execution time but not the probability of success of the exploits.}
over which the one proposed here constitutes an improvement.

\section{The {\secit Combine} Primitive} \label{sec:combine}
\subsection{Predefined Strategies}

We now consider the slightly more general problem where the goal $\frakg$ can be obtained
by predefined strategies.
We call \emph{strategy} a group of actions
that must be executed in a specific order.
The strategies are a way to incorporate the expert knowledge 
of the attacker in the planning system 
(cf. the opening moves in chess).
This idea has been used in the automation of pentesting tools, see \cite{SarWei08}.

\Figure{grafo2.pdf}{ht}{\linewidth}{grafo2}
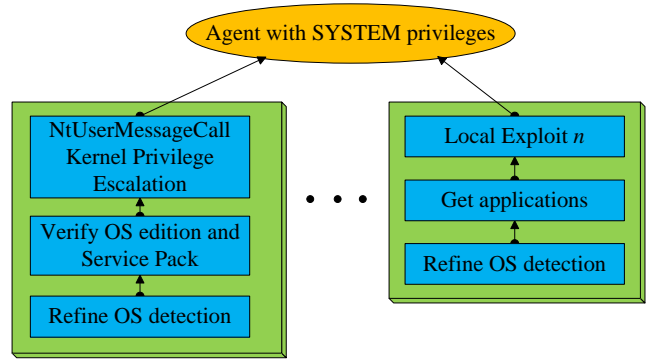
{Multiple strategies for a Local Privilege Escalation.}

For example consider an attacker
who has installed an agent with low privileges on a host $H$ 
running Windows XP, and whose goal
is to obtain \textsc{system} privileges on that host.
The attacker has a set of $n$ predefined strategies to 
perform this privilege escalation (see Fig.~\ref{fig:grafo2}).
An example of a strategy is:
refine knowledge of the OS version;
verify that the edition is Home or Professional, with SP2 installed;
get users and groups;
then launch the local exploit 
\Name{Microsoft NtUserMessageCall Kernel Privilege Escalation}
that (ab)uses the vulnerability CVE-2008-1084.
More generally:

\begin{problem}\label{group_action_order}
Let $\frakg$ be a fixed goal, and $\{ G_1, \ldots, G_n  \}$ a set 
of $n$ strategies,
where each strategy $G_k$ is a group of ordered actions.
For a strategy to be successful, all its actions must be successful.
As in Problem \ref{action_order}, the task is to minimize the expected total cost.

%
\end{problem}

In this problem, actions are executed sequentially,
choosing at each step one action from one group, until the goal $\frakg$ is obtained.
Considering only one strategy $G$, we can calculate
its expected cost and probability of success.
Suppose the actions of $G$ are $\{ A_1, \ldots, A_n \}$
and are executed in that order. Then the expected cost (e.g. expected runtime)
of the group $G$ is
$$
T_G = t_1 + p_1 \, t_2 + p_1 \, p_2 \, t_3 + \ldots
+ p_1 \, p_2 \ldots p_{n-1} \, t_n
$$
and, since all the actions must be successful,
the probability of success of the group is simply
$ P_G = p_1 \, p_2 \ldots p_n $.

\begin{proposition}\label{prop_group_action_order}
A solution to this problem is to sort the 
strategies according to the 
coefficient $T_G / P_G$ (smallest value first), and execute them in that order.
For each strategy group, execute the actions until
an action fails or all the actions are successful.
\end{proposition}

\begin{proof}
In this problem, an attack plan could involve choosing actions
from different groups without completing all the actions of each group.
But it is clear that this cannot happen in an optimal 
plan.\footnote{Suppose that there are only two groups $G_A$ and $G_B$,
whose actions are $\{ A_1, \ldots, A_s \}$ and $\{ B_1, \ldots, B_t \}$ respectively.
Suppose that in the optimal plan $A_s$ precedes $B_t$.
Suppose also that the execution of an action $B_j \neq B_t$ precedes the 
execution of $A_s$.
Executing $B_j$ will not result in success (that requires executing $B_t$ as well),
and it will delay the execution of $A_s$ by 
the expected running time of $B_j$.
Thus to minimize the expected total running time, a better solution 
can be obtained by executing $B_j$ after the execution of $A_s$.
This contradiction shows that all the actions of $G_B$ must be executed after $A_s$
in an optimal solution.
This argument can be easily extended to any number of groups.
}

So an optimal attack plan consists in choosing a group
and executing all the actions of that group.
Since the actions of each group $G$ are executed one after the other,
they can be considered as a single action with probability $P_G$ and expected time $T_G$.
Using the {\em choose} primitive, it follows that groups should be ordered
according to the coefficients $T_G / P_G$.
\end{proof}

\subsection{Multiple Groups of Actions} \label{sec:two-layers-tree}

We extend the previous problem to consider groups of actions
bounded by an AND relation (all the actions of the group must be
successful in order to obtain the result $\frakg$),
but where the order of the actions
is not specified.
The difference with Problem \ref{group_action_order} is that now
we must determine the order of execution within each group.

\Figure{grafo3.pdf}{ht}{\linewidth}{grafo3}
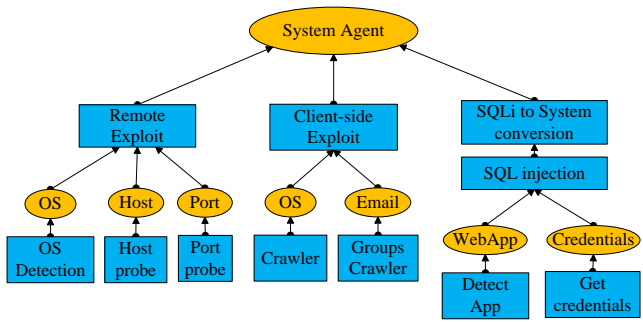
{Probabilistic attack tree (with two layers).}

Fig.~\ref{fig:grafo3} shows an example of this situation. A System Agent
can be installed by using a Remote exploit, a Client-side exploit or 
a SQL injection in a web application.
Each of these actions has requirements represented as assets, which can be
fulfilled by the actions represented on the second layer.
For example, before executing the Remote exploit, the attacker
must run a Host probe (to verify connectivity with the target host),
Port probe (to verify that the target port of the exploit is open),
and an OS Detection module (to verify the OS of the target host).

\begin{problem}\label{free_group_action_order}
Same as Problem \ref{group_action_order}, except that we have
$n$ groups $\{ G_1, \ldots, G_n  \}$ of unordered actions.
If all the actions in a group are successful,
the group provides the result $\frakg$.
\end{problem}


\begin{proposition}\label{order_and_group}
Let $G = \{ A_1, \ldots, A_n \}$ be a group of actions
bounded by an AND relation.
To minimize the expected total cost, the actions must be ordered
according to the coefficient $t_k / (1 - p_k)$.
\end{proposition}

\begin{proof}
If the actions are executed in the order $A_1, \ldots, A_n$, then 
the expected cost is
\begin{align}\label{exp_time_and}
T_G = t_1 + p_1 \, t_2 +  
\ldots + p_1 \, p_2 \ldots p_{n-1} \, t_n
\end{align}
This expression is very similar
to equation \eqref{exp_time_or}.
The only difference is that costs are multiplied by $p_k$ instead of $\overline{p_k}$.
So in this case, the optimal solution is to order the actions according 
to the coefficient $t_k / \overline{p_k} = t_k / (1 - p_k)$. 
\end{proof}

Intuitively the actions that have higher probability of failure 
have higher priority, since a failure ends the execution of the group.
The coefficient $t_k / (1 - p_k)$ represents a trade-off between
cost (time) and probability of failure.

Wrapping up the previous results, to solve Problem \ref{free_group_action_order},
first order the actions in each group according to the coefficient $t / (1-p)$ in increasing order.
Then calculate for each group $G$ the values $T_G$ and $P_G$.
Order the groups according to the 
coefficient $T_G / P_G$, and select them in that order.
For each group, execute the actions until
an action fails or all the actions are successful.

We call it the {\em combine} primitive, because it tells you
how to combine a group of actions and consider them
(for planning purposes) as a single action with probability
of success $P_G$ and expected running time $T_G$.

\section{Using the Primitives in an Attack Tree}\label{sec:dynamic}

We apply below the {\em choose} and the {\em combine} primitives
to a probabilistic attack tree, where the nodes are bounded 
by AND relations and OR relations.
The tree is composed of two types of nodes, distributed 
in alternating layers of asset nodes and action nodes (see Fig.~\ref{fig:grafo4}).

\Figure{grafo4.pdf}{ht}{\linewidth}{grafo4}
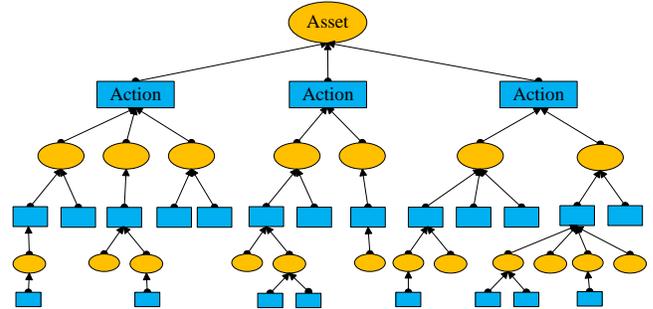
{Attack tree with alternating layers of Assets and Actions.}

An \emph{asset node} is connected by an OR relation to all the actions
that provide this asset: for example, an Agent asset
is connected to the Exploit actions that may install an agent on the target host.

An \emph{action node} is connected by an AND relation to its requirements:
for example, the local exploit 
\Name{Microsoft NtUserMessageCall Kernel Privilege Escalation}
requires an agent asset (with low level privileges) on the target host $H$,
and a Windows XP OS asset for $H$.

The proposed solution is obtained by composing the primitives from previous sections.
In the AND-OR tree, 
the leaves that are bounded by an AND relation 
can be considered as a single node. 
In effect, using the {\em combine} primitive, 
that group $G$ can be considered
as a single action with compound probability of success $P_G$ and execution time $T_G$.

The leaves that are bounded by an OR relation can also
be (temporarily) considered as a single node.
In effect, in an optimal solution, the node that minimizes the $t/p$ coefficient
will be executed first (using the {\em choose} primitive), 
and be considered as the cost of the group in a single step plan.

By iteratively reducing groups of nodes, we build a single path of execution
that minimizes the expected cost.
After executing a step of the plan, the costs may be modified and the shape of the graph may vary.
Since the planning algorithm is very efficient, we can replan after each execution
and build a new path of execution.
We are assured that before each execution, the proposed attack plan is optimal
given the current environment knowledge.

\subsection{Constructing the Tree} \label{sec:tree-construction}

We briefly describe how to construct a tree beginning with an agent asset
(e.g. the objective is to install an agent on a fixed machine). 
Taking this goal as root of the tree, we recursively add the actions that 
can complete the assets that appear in the tree,
and we add the assets required by each action.

To ensure that the result is a tree and not a DAG, we make an additional 
independence assumption: the assets required by each action are considered
as independent (i.e. if an asset is required by two different actions,
it will appear twice in the tree).

That way we obtain an AND-OR tree with alternating layers of asset nodes 
and action nodes (as the one in Fig.~\ref{fig:grafo4}).
The only actions added are Exploits, TCP/UDP Connectivity
checks, and OS Detection modules. These actions don't have as requirements
assets that have already appeared in the tree, in particular 
the tree only has one agent asset (the root node of the tree).
So, by construction, we are assured that no loops will appear,
and that the depth of the tree is very limited.

We construct the tree in this top-down fashion,
and as we previously saw, we can solve it bottom-up to obtain
as output the compound probability of success and the expected running time
of obtaining the goal agent.

%

\section{The Graph of Distinguished Assets} \label{sec:distinguished-assets}

In this section we use the previous primitives to build an algorithm
for \emph{attack planning} in arbitrary networks,
by making an additional assumption of independence between machines.
First we distinguish a class of assets, namely the assets related with agents.
We refer to them as \emph {distinguished assets}.
At the PDDL level, the predicates associated with the agents are considered
as a separate class.

Planning is done in two different abstraction levels: in the \textbf{first level}, we evaluate the cost of 
compromising one \emph{target} distinguished asset from one fixed \emph{source} distinguished asset.
More concretely, we compute 
the cost and probability of obtaining a target agent given a source agent. At this level, the attack plan 
must not involve a third agent.
The algorithm at the first level is thus to construct the attack tree
and compute an attack plan as described in Section \ref{sec:dynamic}.

At the \textbf{second level}, we build a directed graph $\cG = (\cV, \cE)$
where the nodes are distinguished assets
(in our scenario, the hosts in the target network where we may install agents),
and the edges are labeled with the compound probability and expected time
obtained at the first level.
Given this graph, an initial asset $s \in \cV$ (the local agent of the attacker)
and a final asset $g \in \cV$ (the goal of the attack), 
we now describe two algorithms to find a path that 
approximates the minimal expected time of obtaining the goal $g$.

The first algorithm is a modification of Floyd-Warshall's algorithm to find
shortest paths in a weighted graph.
Let $M = | \cV | $ be the number of machines in the target network.
By executing $M^2$ times the first level procedure, we obtain two functions:
the first is $Prob(i,j)$ which returns the compound probability
of obtaining node $j$ from node $i$ (without intermediary hops), 
or $0$ if that is not possible in the target network;
the second is $Time(i,j)$ which returns the expected time
of obtaining node $j$ directly from node $i$, or $+\infty$ if that is not possible.
The procedure is described in Algorithm~\ref{floyd-warshall}.

\begin{algorithm}[h]
\caption{Modified Floyd-Warshall}
\label{floyd-warshall}
\algsetup{indent=1.5em}
\fontsize{9}{12}\selectfont
\begin{algorithmic}
\STATE $P[i,j] \gets Prob(i,j) \quad \forall \; 1 \leq i,j \leq M $
\STATE $T[i,j] \gets Time(i,j) \quad \forall \; 1 \leq i,j \leq M $

\FOR{$k = 1$ to $M$}
  \FOR{$i = 1$ to $M$}
    \FOR{$j = 1$ to $M$}
      \STATE $ T' \gets T[i,k]  +  P[i,k] \times T[k,j] $
      \STATE $ P' \gets P[i,k] \times P[k,j] $
      \IF{${T'}/{P'} \, < \, T[i,j] / P[i,j]$}
        \STATE $T[i,j] \gets T' $
        \STATE $P[i,j] \gets P' $
      \ENDIF
    \ENDFOR
  \ENDFOR
\ENDFOR
\RETURN $ \langle T, P \rangle $
\end{algorithmic}
\end{algorithm}

When the execution of this algorithm finishes, for each $i,j$ 
the matrices contain the compound probability $P[i,j]$ and
the expected time $T[i,j]$ of obtaining the node $j$
starting from the node $i$. This holds in particular when $i = s$ (the source
of the attack) and $j = g$ (the goal of the attack).
The attack path is reconstructed just as in the classical Floyd-Warshall algorithm.

In a similar fashion, Dijkstra's shortest path algorithm can be modified to use
the {\em choose} and {\em combine} primitives. See the description of Algorithm \ref{dijkstra}.

{
\begin{algorithm}[h]
\caption{Modified Dijkstra's algorithm}
\label{dijkstra}
\algsetup{indent=1.5em}
\fontsize{9}{12}\selectfont
\begin{algorithmic}

\STATE $ T[s] = 0,\; P[s] = 1 $
\STATE $ T[v] = +\infty,\; P[v] = 0 \quad \forall v \in \cV, v \neq s $
\STATE $ S \gets \emptyset $
\STATE $ Q \gets \cV $ (where $Q$ is a priority queue)
\WHILE{$ Q \neq \emptyset $}
  \STATE $ u \gets \arg \min_{x \in Q} T[x]/P[x]$
  \STATE $ Q \gets Q \backslash \{ u \}, \; S \gets S \cup \{ u \} $
  \FORALL{ $v \in \cV \backslash S $ adjacent to $u$ }
    \STATE $ T' = T[u] + P[u] \times Time(u,v) $
    \STATE $ P' = P[u] \times Prob(u,v) $
    \IF{${T'}/{P'} < T[v] / P[v] $}
      \STATE $ T[v] \gets T' $
      \STATE $ P[v] \gets P' $
    \ENDIF
  \ENDFOR
\ENDWHILE
\RETURN $ \langle T, P \rangle $
\end{algorithmic}
\end{algorithm}
}

When execution finishes, the matrices contain the compound probability $P[v]$ and
the expected time $T[v]$ of obtaining the node $v$
starting from the node $s$.
Using the modified Dijkstra's algorithm has the advantage that its complexity
is $\calO( M ^ 2)$ instead of $\calO( M ^3 ) $ for Floyd-Warshall.
Let $n$ be the number of actions that appear in the attach trees,
this gives us that the complexity of the complete planning solution is
$\calO( M^2 \cdot n \log n + M^2 ) = \calO( M^2 \cdot n \log n )$.

\section{Our implementation} \label{sec:experiments}

We have developed a proof-of-concept implementation of these ideas
in the Python language. This planner takes as input a description of the scenario
in the PPDDL language, an extension of PDDL for expressing probabilistic effects \cite{YouLit04}.

Our main objective was to build a probabilistic planner
able to solve scenarios with 500 machines, which was the limit reached with classical
(deterministic) planning solutions in \cite{LucSarRic10}.
Additionally we wanted to tame memory complexity, which was the limiting factor.
The planner was integrated with the pentesting framework Core Impact, 
using the procedures previously developed for the work \cite{LucSarRic10}.
The architecture of this solution is described in Fig.~\ref{fig:architecture}.

\Figure{architecture.pdf}{ht}{\linewidth}{architecture}
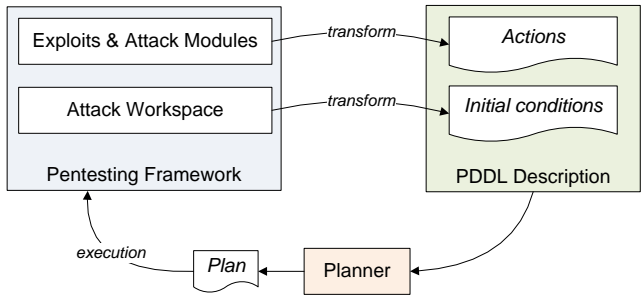
{Architecture of our solution.}

This planner solves the planning problem by breaking it into two levels
as described in Section~\ref{sec:distinguished-assets}.
On the higher level, a graph representation of \emph{goal} objects is built. 
More concretely, there is a distinguished node for each host.
The directed edges in this graph are obtained by carrying out the tree procedure 
described in Section \ref{sec:tree-construction},
obtaining a value for the probability and the cost of obtaining the predicate represented 
by the target node, when the predicate represented by the source node is true. 

The final plan can then be determined by using the modified versions
of Dijkstra and Floyd-Warshall algorithms.
The figures that follow show the planner running time using
the modified Dijkstra's algorithm.

\subsection{Testing and Performance}

The experiments were run on a machine with an Intel Core2 Duo CPU at 2.4 GHz and 8 GB of RAM.
We focused our performance evaluation on the number of machines $M$ in the attacked network.
We generated a network consisting of five subnets with varying number of machines,
all joined to one main network to which the attacker initially has access.

\Figure{plot_mem_abs.pdf}{t}{\linewidth}{memabs}
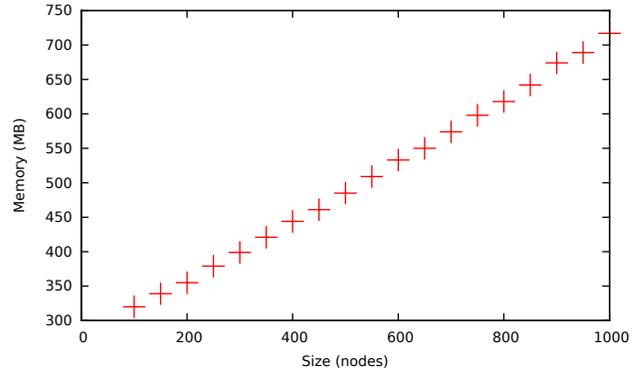
{Memory consumption vs number of machines.}

\Figure{plot_time_abs.pdf}{t}{\linewidth}{timeabs}
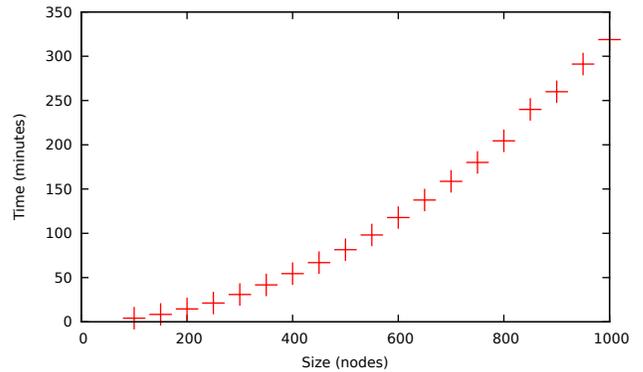
{Solver runtime vs number of machines.}

\Figure{plot_relative.pdf}{t}{\linewidth}{plotrel}
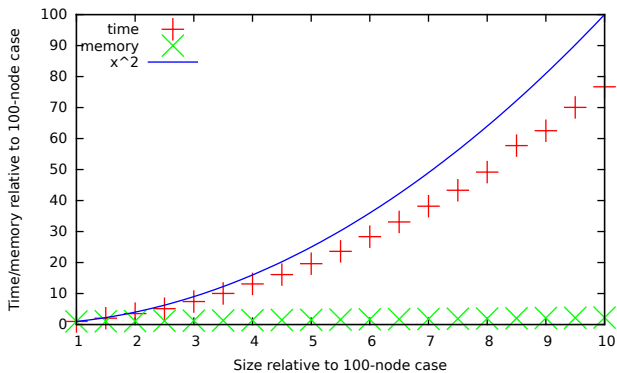
{Time and memory relatives to the 100 machines case.}

Fig.~\ref{fig:memabs} shows the memory consumption of this planning solution, 
which clearly grows linearly with $M$.
Our current implementation manages to push the network size limit up to 1000 machines, and brings memory 
consumption under control.\footnote{By contrast, in \cite{LucSarRic10} the hard limit was memory:
in scenarios with 500 machines we ran out of memory in a computer with 8 GB of RAM.
The memory consumption growth was clearly exponential, for instance
400 machines used 4 GB of RAM. This was difficult to scale up. }
For $M = 1000$, we are using less than 1 GB of RAM,
with a planner completely written in Python (not optimized in terms of memory consumption).

Fig.~\ref{fig:timeabs} shows the growth of solver running time, which seems clearly quadratic,
whereas in \cite{LucSarRic10} the growth was exponential.
It should be noted however that, comparing only up to 500 machines, 
running times are slightly worse than those of the solution based 
on deterministic planners.
This can be improved: since our planner is written in Python, a reasonable implementation in C 
of the more CPU intensive loops should allow us to lower significantly the running time. 

And of course we added a notion of probability of success that wasn't present before.
As a comparison, in another approach that accounts for the uncertainty about
the attacker's actions \cite{SarBufHof11},
the authors use off-the-shelf solvers, managing to solve scenarios with up to 7 machines -- 
and are thus still far from the network sizes reached here.


Both curves are compared in Fig.~\ref{fig:plotrel} showing the quadratic growth of solver runtime.
In the testing scenarios, the nodes are fully connected,
so we have to solve a quadratic number of attack trees.
This figure also confirms in practice the computed complexity.

An interesting characteristic of the solution proposed is that
it is inherently parallelizable. 
The main workload are the $M^2$ executions of the \emph{first level} procedure 
of Section~\ref{sec:distinguished-assets}.
This could be easily distributed between CPUs or GPUs to obtain a faster planner.
Another possible improvement is to run the planner ``in the cloud'' 
with the possibility of adding processors on demand.

\section{Related Work}

Early work on attack graph solving relied on model checking techniques \cite{JhaSheWin02,SheHaiJha02},
with their inherent scalability restrictions; or on monotonicity assumptions \cite{AmmWijKau02,NoeEldJaj09,NoeJaj05} 
that are not able to express situations in which compromised resources are lost due to crashes, 
detection or other unforeseen circumstances.

The first application of planning techniques and PDDL solving for the security realm was \cite{BodGohHaiHar05},
however this application was not focused on finding actual attack paths or driving penetration testing tools. 
In \cite{GhoGho09} attack paths are generated from PDDL description of networks, hosts and exploits, 
although the scenarios studied do not cover realistic scales. 
Previous work by the authors \cite{LucSarRic10} addresses this limitation by solving scenarios with up 500 machines, and feeding the generated attack plans to guide a penetration testing tool. However, this work does not include probabilistic considerations. Recent work \cite{ElsKohMen11} also manages to provide attack paths to a penetration testing tool, in this case the Metasploit Framework, but again does not include probabilistic considerations.

Previous work by one of the authors \cite{SarBufHof11} takes into account the uncertainty about the result of the attacker's actions. This POMDP-based model also accounts for the uncertainty about the target network, addressing information gathering as an integral part of the attack, and providing a comprehensive notion of attack planning under uncertainty. 
However, as previously stated, this solution does not scale to medium or large real-life networks.

\section{Summary and Future Work}

We have shown in this paper an extension of
established \emph{attack graphs} models, 
that incorporates probabilistic effects,
and numerical effects (e.g. the expected running time of the actions).
This model is more realistic than the deterministic setting,
but introduces additional difficulties to the planning problem.
We have demonstrated that under certain assumptions,
an efficient algorithm exists that provides optimal attack plans
with computational complexity $\calO ( n \log n) $, where 
$n$ is the number of actions and assets in the case of an attack tree
(between two fixed hosts),
and $\calO ( M^{2} \cdot n \log n) $ where $M$ is the number of machines
in the case of a network scenario.

Over the last years, the difficulties that arose in our research
in \emph{attack planning} were related to the exponential nature of planning
algorithms (especially in the probabilistic setting), and our efforts
were directed toward the aggregation of nodes and simplification of the
graphs, in order to tame the size and complexity of the problem.
Having a very efficient algorithm in our toolbox gives us a new direction of research:
to refine the model, and break down the actions in smaller parts,
without fear of producing an unsolvable problem.

A future step in this research is thus to analyze and divide the exploits
into basic components. 
This separation gives a better probability distribution of the
exploit execution. For example, the
\Name{Debian OpenSSL Predictable Random Number Generation Exploit}
-- which exploits the vulnerability CVE-2008-0166 reported by Luciano Bello --
brute forces the 32,767 possible keys.
Each brute forcing iteration can be considered as a basic action, 
and be inserted independently in the attack plan.
Since the keys depend on the Process ID (PID), some keys are more probable than 
others.\footnote{The OpenSSL keys generated in vulnerable Debians only depend on the PID. 
Since Secure Shell usually generates the key in a new installation,
PIDs between 2,000 and 5,000 are more probable than the others.}
So the planner can launch the \Name{Debian OpenSSL PRNG} exploit, 
execute brute forcing iterations for the more probable keys, 
switch to others exploits and come to back to the Debian PRNG exploit if the others failed.
This finer level of control over the exploit execution should produce
significant gains in the total execution time of the attack.

Other research directions in which we are currently working are
to consider actions with multidimensional numeric effects
(e.g. to minimize the expected running time and generated network traffic 
\emph{simultaneously});
and to extend the algorithm to solve probabilistic attack planning in 
Directed Acyclic Graphs (DAG) instead of trees.
In this setting, an asset may influence the execution of several actions.
This relaxes the independence assumption of Sections \ref{sec:two-layers-tree} 
and \ref{sec:dynamic}.
Although finding a general algorithm that scales to the network sizes that we consider 
here seems a difficult task, we believe that efficient 
algorithms specifically designed for network attacks scenarios can be found.

\subsection*{Acknowledgments}

Thanks to Ariel Futoransky and Ariel Waissbein
for their contributions and insightful discussions.


%

\end{document}